\documentclass[prd,onecolumn]{revtex4}
\usepackage{graphicx}
\begin{document}
\title{LATOR Covariance Analysis}
\date{May 26, 2005}
\author{Joseph E. Plowman}
\email{plowman@physics.montana.edu}
\author{Ronald W. Hellings}
\email{hellings@physics.montana.edu}
\affiliation{Department of Physics \\ Montana State University \\ Bozeman, MT 59717}
\begin{abstract}
We present results from a covariance study for the proposed Laser Astrometric Test of Relativity (LATOR) mission. This mission would send two laser-transmitter spacecraft behind the Sun and measure the relative gravitational light bending of their signals using a hundred-meter-baseline optical interferometer to be constructed on the International Space Station. We assume that each spacecraft is equipped with a $ < 1.9 \times 10^{-13} \mathrm{m} \mathrm{s}^2 \mathrm{Hz}^{-1/2} $ drag-free system and assume approximately one year of data. We conclude that the observations allow a simultaneous determination of the orbit parameters of the spacecraft and of the Parametrized Post-Newtonian (PPN) parameter $\gamma$ with an uncertainty of $2.7 \times 10^{-9}$. We also find a $5 \times 10^{-9}$ determination of the solar quadrupole moment, $J_2$, as well as the first measurement of the second-order post-PPN parameter $\delta$ to an accuracy of about $10^{-3}$.
\end{abstract}
\maketitle
\section{Introduction}
In this paper, we present results of a covariance study for the
proposed Laser Astrometric Test of Relativity (LATOR) mission (see \citet{0264-9381-21-12-001}). The
goal of the LATOR Mission is to effect a precise measurement of
the gravitational bending of light for laser signals passing near the Sun and thereby determine several Parametrized
Post-Newtonian (PPN) parameters of relativistic gravity with an
accuracy many orders of magnitude better than present best estimates.

\begin{figure}
\includegraphics[width=6in,keepaspectratio=true]{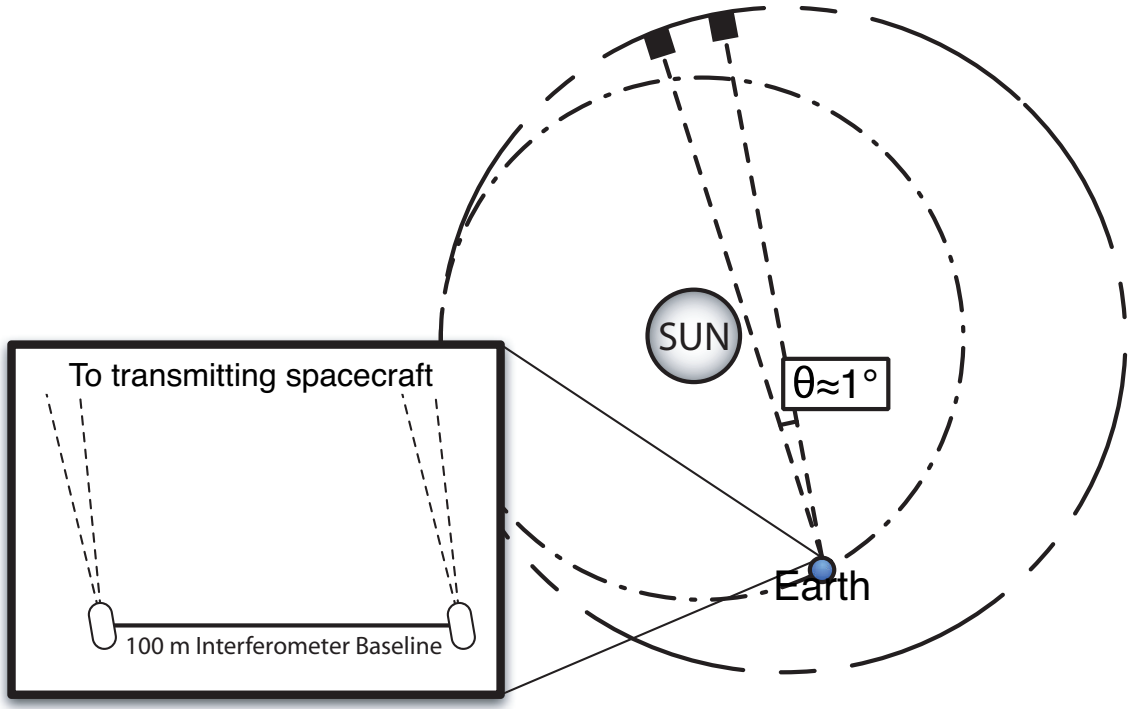}
\caption{\label{latorf1} Diagram of LATOR geometry: The transmitting spacecraft are on far side of the Sun from Earth; the receiving spacecraft are in low Earth orbit aboard the ISS}
\end{figure}

\begin{figure}
\includegraphics[width=6in,keepaspectratio=true]{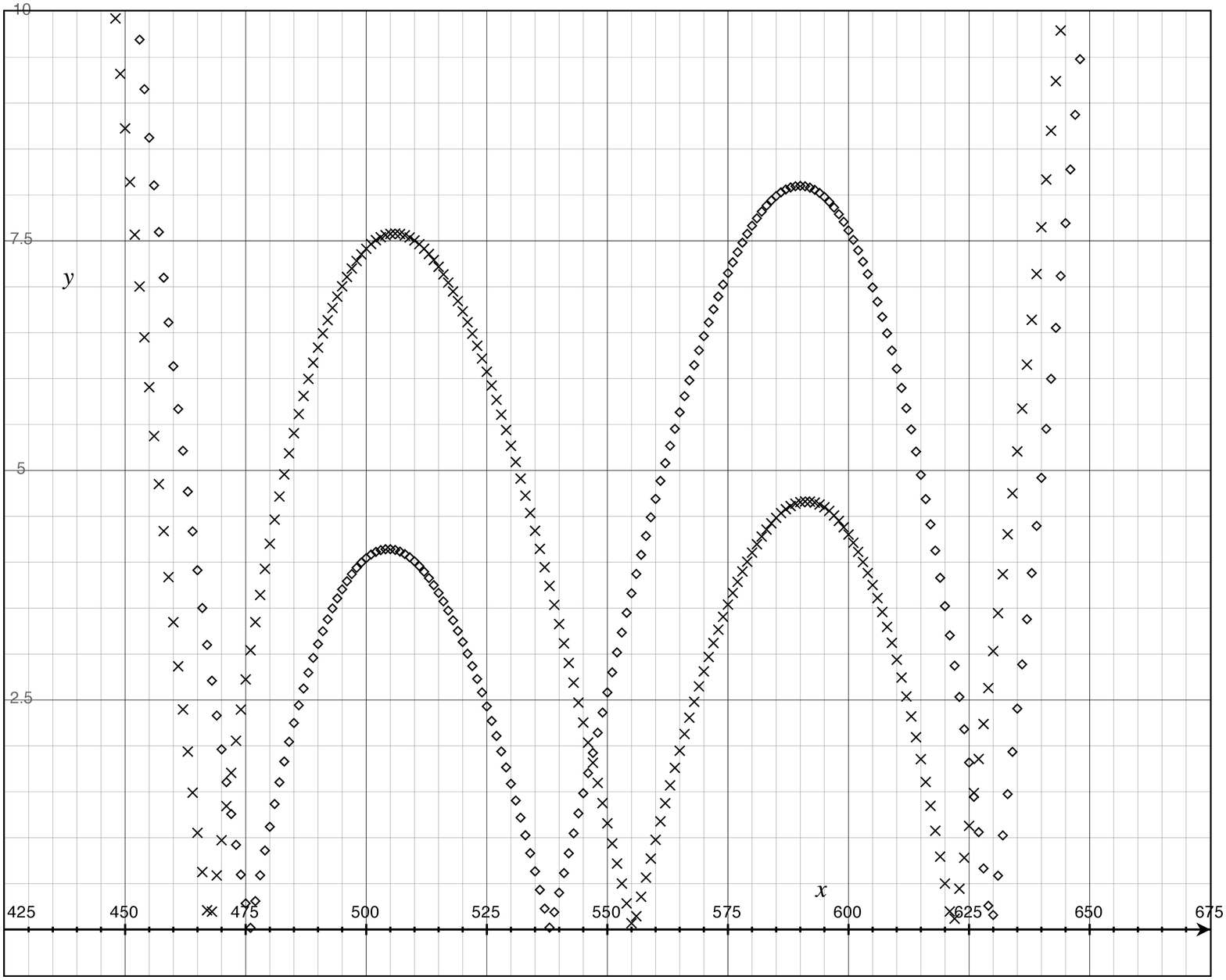}
\caption{\label{orbitplot} Plot of impact parameter vs. time for orbits of spacecraft 1 (crosses) and spacecraft 2 (diamonds), with 1 day sampling interval. Time axis is in days, vertical (impact parameter) axis is in units of $R_{SUN}$.}
\end{figure}

The LATOR instrument consists of two Sun-orbiting
laser transmitter spacecraft in eccentric ecliptic orbits with a
3/2 period resonance with the Earth's orbit.  The two spacecraft
are separated by about 1 degree, as seen from the Earth.  The
angular position of each spacecraft is measured using a
100m-baseline optical interferometer, proposed for the
Earth-orbiting International Space Station (ISS). Figure
\ref{latorf1} shows the LATOR mission geometry.  The coherent
laser signals from each spacecraft will allow the position of each
to be determined with an accuracy of 0.1 picoradians (20
nanoarcseconds), relative to the reference frame defined by the
interferometer baseline. The simultaneous measurement of two
spacecraft enables a relative angle to be measured, eliminating
the need to tie the interferometer frame to inertial space.
As the lines-of-sight from the spacecraft to the Earth pass close
to the Sun, the gravitational bending of light will affect the
inner spacecraft's apparent position more than it does the outer
spacecraft, causing the inter-spacecraft angle, as seen from Earth, to differ from its value in the absence of gravitational effects. Measurements of this inter-spacecraft angle may thus detect the gravitational bending of light and verify gravitational theories that predict it.
Many measurements of this angle will be made as, over the course of a year, the spacecraft will make three such passes behind the Sun, giving detailed information about how the bending of light changes depending on which part of the solar gravitational field the light passes through (see figure \ref{orbitplot}). For more information on the design of LATOR, see \citet{Turyshev:2004vn,Turyshev:2004ga}.

One of the problems, of course, in measuring the gravitational
deflection of the laser signals from LATOR is to know what the
apparent angle between the two spacecraft would be without the
gravitational bending present.  Two methods have been proposed to
provide this knowledge.  One is to have a microwave (or optical) signal sent
between the two spacecraft, determining their separation to an
accuracy of $\sim 1$ cm.  The other is to equip each LATOR
spacecraft with a drag-free system, so that the orbit of each is
very close to a purely gravitational and noise-free trajectory.
The spacecraft-to-spacecraft distance can then be inferred by
accurate tracking and orbit-modeling over the time of the
mission. As we will discuss in section \ref{ipsec}, the
spacecraft-to-spacecraft link leaves one important problem
unsolved, that of determining the impact parameter.  However, as we show by means of this
covariance study, the drag-free system allows orbit modeling
using the primary LATOR data to simultaneously solve the
spacecraft-to-spacecraft distance and the impact parameter
problems.

\section{LATOR Science Goals}
As was shown in the early 1970's, most theories of gravity are
metric theories in which the dynamics of neutral matter are
determined by a metric, which is itself determined by the
neighboring mass distribution. \citet{1972ApJ...177..757W}
developed an expansion of the metric that has a tunable set of
numerical parameters which allow it to describe a very broad range
of metric theories of gravity.  This Parameterized Post-Newtonian
(PPN) expansion is useful for experimental tests of gravity
because the task of fitting the model to the data is carried out
by simply adjusting the numerical values of these tunable
parameters so that the model best fits the data. For the LATOR
experiment, the version of the PPN metric required is:
\begin{eqnarray} 
g_{00} & = & -1+2 \frac{GM}{r}(1-J_2\frac{R^2}{r^2}\frac{3 \cos^2 \theta - 1}{2})-2\beta \frac{G^{2}M^{2}}{r^{2}} \label{eq:g0k1}\\
g_{lm} & = & \delta_{lm}(1+2\gamma \frac{GM}{r}[1-J_2\frac{R^2}{r^2}\frac{3 \cos^2 \theta - 1}{2}]+\frac{3}{2} \delta \frac{G^{2}M^{2}}{r^{2}}),\label{eq:glk1}
\end{eqnarray}
where $M$ is the mass of the Sun, $R$ is the solar radius, $r$ and $\theta$ are polar coordinates, $J_2$ is the solar quadrupole moment, and $\gamma$, $\beta$ and $\delta$ are PPN parameters.
While the full PPN metric contains other terms as well, this
version contains the terms that contribute significantly to
the bending of light that is observable by LATOR. There is a frame-dragging term that scales with the solar rotational angular momentum, $J_z$, and which should have an effect similar in size to that of other terms observable by LATOR (see \citet{1983PhRvD..28.3007R}
). This term is not included in our model, but should not be difficult to add at a later date.
Table~\ref{ppnsig} contains a description of the effects of each
parameter in the above metric, along with its current best
estimate and theoretical value in Einstein's General Relativity (GR).
\begin{table}[h]
\begin{tabular}[c]{|c|p{4.7cm}|p{4.7cm}|}
\hline
Parameter & Significance & Current best estimate \\
\hline
\hline
$\gamma$ & Space-time curvature per unit rest mass. & $ 1 + (2.1 \pm 2.3)\times 10^{-5} $\cite{Bertotti:2003rm}. Equal to one in GR \\
\hline
$\beta$ & Space curvature per unit gravitational self-energy & $1+(0.9\pm 1.1)\times 10^{-4}$\cite{2004PhRvL..93z1101W}. Equal to one in GR  \\
\hline
$\delta$ & Time curvature per unit gravitational self-energy & Unmeasured. Equal to one in GR \\
\hline
$J_2$ & Solar quadrupole moment parameter. & Never measured directly. Estimated  from solar models to be about $10^{-7}$. \\
\hline
\end{tabular}
\caption{\label{ppnsig} Significance and best estimates for parameters in PPN metric}
\end{table}

\begin{figure}
\includegraphics[width=6in,keepaspectratio=true]{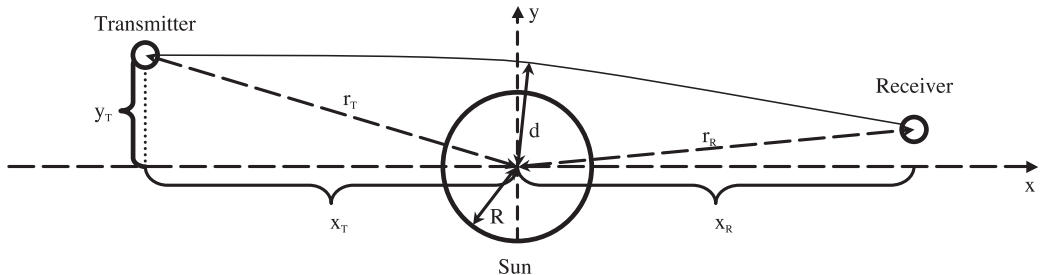}
\caption{\label{tofeqdiagram} Diagram of coordinate system used for equation \ref{tofeq}.}
\end{figure}

The accuracy of an interferometer angular measurement of gravitational
deflection is related to the accuracy of the measurement of the
differential time-of-arrival of a phase front in the two detectors
of the interferometer.  The relation is
\begin{displaymath} 
\Delta\theta=\frac {c\Delta\tau} {L}
\end{displaymath}
where $\Delta\theta$ is the accuracy of the angular measurement in
radians, $\Delta\tau$ is the accuracy of the timing measurement in
seconds, and $L$ is the interferometer baseline in meters. LATOR's 20 nanoarcsecond (or $10^{-13}$ radian) angular accuracy thus corresponds to
$3.33\times10^{-20}$ s (or about 20 $\mu$cycles) accuracy in
timing measurement at each element of the interferometer.
To second order in $GM/r$, the time delay derived from the
metric of eqs. \ref{eq:g0k1} and \ref{eq:glk1} may be written (see
\citet{1983PhRvD..28.3007R})
\begin{eqnarray}
\frac{\tau_{TR}}{R} & = &\frac{x_R-x_T}{R}(1-\frac{GM}{R}\frac{R}{r_T}+\frac{1}{2}(1+\gamma^2)\frac{(GM)^2}{R^2}\frac{R^2}{y_T^2}((1-\frac{x_T}{r_T})^2+2\frac{x_R x_T-y_T^2}{r_R r_T}) \nonumber\\
& & {}+\frac{1}{2}(2\beta-1)\frac{(GM)^2}{R^2}\frac{R^2}{r_T^2})+((1+\gamma)\frac{GM}{R}(1-\frac{GM}{R}\frac{R}{r_T}))\ln\frac{x_R+r_R}{x_T+r_T} \nonumber\\
& & {}+(2(1+\gamma)-\beta+\frac{3}{4}\delta )\frac{(GM)^2}{R^2}\frac{R}{y_T}(\arctan\frac{x_R}{y_T}-\arctan\frac{x_T}{y_T})+ \nonumber\\
& & {}+(1+\gamma)^2\frac{(GM)^2}{R^2}\frac{R}{y_T}\frac{r_T}{y_T}(1-\frac{r_R}{r_T})(\frac{x_T}{r_T}+\frac{x_R}{r_R}-1)+\frac{1}{2}(1+\gamma) J_2 \frac{GM}{R}\frac{R^2}{y_T^2}\label{tofeq}
\end{eqnarray}
Here, it is assumed that the $x$ axis is tangent to the photon
trajectory at the transmitter, the origin is at the position of the
source mass (the Sun), and that the Sun, transmitter and receiver are
in the $z=0$ plane. $R$ is the radius of the Sun, $x_R$ and $r_R$
are the receiver $x$-coordinate and distance from the origin,
$x_T$ and $y_T$ are the transmitter $x$- and $y$-coordinates, and
$r_T$ is the transmitter distance from the origin (see figure \ref{tofeqdiagram}).

The sizes of each term in eq. \ref{tofeq} for photons
travelling 2 AU and just grazing the limb of the Sun are given in
table \ref{sttab}, also taken from \citet{1983PhRvD..28.3007R}.
\begin{table}[h]
\begin{tabular}{|c|c|}
\hline
Term: & Size of term in seconds: \\
\hline
\hline
$4\frac{r}{R}$ & $2000$ \\
\hline
$2(1+\gamma)\frac{GM}{R}\ln (4(\frac{r}{R})^2)$ & $2.4 \times 10^{-4}$ \\
\hline
$4\frac{GM}{R}$ & $2.0\times 10^{-5}$ \\
\hline
$4(1+\gamma)^2(\frac{GM}{R})^2 \frac{r}{R}$ & $3.6\times 10^{-8}$ \\
\hline
$ 2(1+\gamma)J_2 \frac{GM}{R} $ & $ 4.9 \times 10^{-10} $ \\
\hline
$4[2(1+\gamma)-\beta+\frac{3}{4}\delta](\frac{GM}{R})^2\arctan(\frac{r}{R})$ & $ 2.5 \times 10^{-10} $ \\
\hline
$2(1+\gamma)(\frac{GM}{R})^2 \frac{R}{r} \ln (4(\frac{r}{R})^2)$ & $2.4 \times 10^{-12} $\\
\hline
$2(2\beta-1)(\frac{GM}{R})^2 \frac{R}{r}$ & $9.7\times 10^{-14}$ \\
\hline
\end{tabular}
\caption{\label{sttab} Size of terms in time-of-flight expression for 2 AU light
path which grazes the limb of the Sun. PPN parameters are set to
their values in GR, and $J_2$ and $J_z$ are set to their values in Dicke's model
of the Sun.}
\end{table}
The sizes of these terms should be compared with the timing
accuracy expected for the LATOR interferometer measurement,
remembering that the actual differential times-of-flight
measurements for the two LATOR signals are a fraction of the
displayed terms in table \ref{sttab}.

The leading term, after the 2000-second geometrical time-of-flight
term, is the term proportional to $1+\gamma$, of size
$2.4\times10^{-4}$, with its second-order correction of size
$3.6\times10^{-8}$.  This is the effect whose measurement will
give PPN parameter $\gamma$ from the LATOR measurements.  The next
largest term, with size $4.9\times10^{-10}$ enables the solar
quadrupole moment, $J_2$, to be measured, while the term of size $2.5\times10^{-10}$
will give a linear combination of $\gamma$, $\beta$, and the
post-post-Newtonian term $\delta$.

The fundamental current incompatibility of General Relativity
and Quantum Mechanics suggests that the PPN parameters should
deviate from their values in GR at some level. The size of the
deviation of the PPN parameters from their value in GR is theory
dependent. \citet{1993PhRvL..70.2217D,1993PhRvD..48.3436D} suggest
that for some scalar-tensor theories of gravity the deviation in
the PPN parameter $\gamma$ occurs at the level $1-\gamma \sim
10^{-7}$, significantly smaller than the uncertainty in the current best estimate for $\gamma$, at $\sim 10^{-5}$ (see \citet{Bertotti:2003rm}). The primary goal of the LATOR mission is measurement of
the PPN parameter $\gamma$ to an accuracy of $10^{-9}$.
Secondarily, LATOR will provide the first measurement of a
post-post-Newtonian parameter, $\delta$, and a measurement of the
solar quadrupole moment parameter $J_2$, which will provide
information about the solar interior.

\section{Covariance Study}
We model the LATOR instrument in the solar system and use a linear
least squares covariance analysis to solve simultaneously for the
uncertainties in the PPN parameters and orbit parameters. A
Newtonian model is used for the transmitter orbits, and time
delays for the interferometer legs are modeled using eq. \ref{tofeq}. {\it A priori} uncertainties for the orbit parameters are assumed, consistent with typical uncertainties in spacecraft radio tracking.
The transmitter coordinates are evaluated as functions
of the time and of the orbit parameters for each spacecraft, while
receiver coordinates are evaluated as functions of time using a
simple circular heliocentric orbit. A justification for this simple model of the transmitter and receiver orbits is given in appendix \ref{orbjust}. The interferometer is assumed to be inertially oriented. We model two constant body-fixed accelerations for each transmitter, one along the Earth line-of-sight and
one perpendicular to it.  For each assumed data point, the
time-of-flight from each spacecraft to each antenna is calculated
as a function of the time of observation and of the parameters of
the model.  The parameters include the
constants of motion for the transmitter orbit (one set for each
transmitter), the DC accelerations on each spacecraft (one set per
transmitter) and the PPN parameters. These parameters are listed
in table \ref{prms}.  The goal of the covariance analysis is to
estimate the uncertainties with which these parameters will be
determined in the final LATOR mission data analysis.

\begin{table}
\begin{tabular}{|c|p{13cm}|}
\hline
Parameter & Description \\
\hline
\hline
$a_1$, $a_2$ & Semi-major axes for orbits of transmitting spacecraft 1 and 2. \\
\hline
$e_1$, $e_2$ & Eccentricities for orbits of transmitting spacecraft 1 and 2. \\
\hline
$\phi_{01}$, $\phi_{02}$ & Initial phase for orbits of transmitting spacecraft. \\
\hline
$\omega_1$, $\omega_2$ & Angle of semi-major axes of transmitter orbits with respect to x axis. \\
\hline
$a_{\perp 1}$,$a_{\perp 2}$ & Constant acceleration in drag-free system, perpendicular to transmitter-receiver line-of-sight, for transmitting spacecraft. \\
\hline
$a_{\parallel 1}$,$a_{\parallel 2}$ & Constant acceleration in drag-free system, parallel to transmitter-receiver line-of-sight, for transmitting spacecraft. \\
\hline
$\gamma$ & Space-time curvature per unit rest mass. \\
\hline
$\delta$ & Time curvature per unit gravitational self energy. \\
\hline
$J_2$ & Solar quadrupole moment parameter.\\
\hline
\end{tabular}
\caption{\label{prms} Parameters to be solved for in covariance study}
\end{table}

Notably absent from the list of parameters is the parameter
$\beta$. This is because the only significant term in the
time-of-flight equation which contains $\beta$ (see table
\ref{sttab}) is
\begin{displaymath}
4[2(1+\gamma)-\beta+\frac{3}{4}\delta](\frac{GM}{R})^2\arctan(\frac{r}{R})
\end{displaymath}
in which $\beta$, $\delta$ and $\gamma$ are all completely
correlated. Other terms give $\gamma$, but the only other term to
contain $\beta$ is smaller, meaning that $\beta$ would be less well
determined than the combination $\beta+3/4\delta$ and that the
incertainty in both would be that of $\beta$ alone.  However,
since $\beta$ is already known from other solar system experiments (\citet{2004PhRvL..93z1101W}) with an accuracy (see  table \ref{ppnsig}) better than the best LATOR can manage for this term, our analysis assumes that $\beta$ is
already known. The term
\begin{displaymath}\nonumber
4[2(1+\gamma)-\beta+\frac{3}{4}\delta](\frac{GM}{R})^2\arctan(\frac{r}{R})
\end{displaymath}
is therefore used to solve for $\delta$ alone.

At each time sampled, the LATOR signal is formed from
\begin{equation}
S = [\tau_{12}-\tau_{11}]-[\tau_{22}-\tau_{21}]\label{signaleq}
\end{equation}
where each $\tau$ is the time-of-flight for a particular transmitter-receiver pair; so, for example, $\tau_{12}$ would be the time-of-flight between transmitter 1 and receiver 2. At each time, the contribution from the current time step to the
information matrix is formed by calculating the derivatives of $S$
with respect to the parameters of table \ref{prms}.  Referring to
the parameter set collectively as $\vec{x}$, the contribution to
the information matrix at each time $t_i$ is given by
\begin{equation}
\alpha_{jk}(t_i) = \frac{1}{(\sigma)^2}\frac{\partial S(t_i,\vec{x})}{\partial x_j} \frac{\partial S(t_i,\vec{x})}{\partial x_k}\label{fmcon}
\end{equation}
where $\sigma$ is the uncertainty in LATOR's measurement of each
time delay.  In our analysis, $\sigma$ was set to
$3\times10^{-20}$ s, corresponding to the conservative 10 picometer interferometer uncertainty estimate found in LATOR literature (see, for instance, \citet{0264-9381-21-12-001}).

The derivatives in eq. \ref{fmcon} were calculated using analytic
derivatives for PPN parameters, finite differencing for orbit
constants of motion, and finite differencing with a perturbative
Enke integration for the DC accelerations.  Calculations leading
up to the time-of-flight calculation were performed in double
precision. Round-off error was a serious problem for derivatives
with respect to orbit parameters, so we used Ridders'
extrapolative finite differencing algorithm (from \citet{nr}
chapter 5) and performed all calculations after the
time-of-flight calculation in the GNU Multiple Precision
Arithmetic Library (GMP) in order to keep sufficient precision to
solve for uncertainties in all of the variables.

The terms in the information matrix, $\alpha_{jk}(t_i)$, were then
summed over a period of about 1 year, during which the
transmitting spacecraft pass behind the Sun three times as viewed
from Earth (see figure \ref{orbitplot}). We used a 1-minute sampling cadence, but the number of
data points was further reduced because the transmitting
spacecraft are only visible from Earth for half of the ISS
hour-and-a-half orbit, and it may take tens of minutes per orbit
for the receiving interferometer to acquire lock. This was
simulated by adding to the information matrix for only 10 minutes
out of a 90 minute period. The resulting integration has about
50000 time steps. The total information matrix is
\begin{equation}
\alpha_{jk} = \sum_i \frac{1}{(\sigma)^2}\frac{\partial S(t_i,\vec{x})}{\partial x_j} \frac{\partial S(t_i,\vec{x})}{\partial x_k}\label{fm}
\end{equation}

The information matrix was inverted using a Singular Value
Decomposition (SVD), and the diagonal elements of the inverse gave
the standard variations for the corresponding parameters:
\begin{equation}
\sigma_i=\sqrt{\alpha^{-1}_{ii}} \qquad \qquad \textrm{(no sum)}\label{vareq}
\end{equation}

\section{The Impact Parameter Problem}\label{ipsec}
Since the leading $\gamma$ term in the time-of-flight formula is
logarithmic, its uncertainty scales as $1/r$.  We will so far
anticipate the results of our analysis to say that we expect
$\gamma$ to be determined to an accuracy $3 \times 10^{-9}$.  If this is to
be accomplished, then the size of the coefficient of $\gamma$ must
be known correctly to the same accuracy.  For signals passing close
to the Sun, the $1+\gamma$ term in eq. \ref{tofeq} may be approximated as
\begin{equation}\label{approx}
(1+\gamma)\frac{GM}{R}(1-\frac{GM}{R}\frac{R}{r_T})\ln\frac{x_R+r_R}{x_T+r_T}
\approx (1+\gamma)\frac{GM}{R}\ln(4r_Rr_T/d^2)
\end{equation}
where $d$ is the impact parameter.  Thus, if the coefficient of
$\gamma$ is to be known to $3 \times 10^{9}$, the impact
parameter must be known to a part in $10^9$.  For paths passing
near the solar limb, the impact parameter is $\sim10^{9}$ m,
meaning that it must be known to within a meter.  How is this to
be done?

First, we point out that the spacecraft-to-spacecraft link that
could be used to give the base of the narrow triangle linking the
two spacecraft and the interferometer provides no information on
how that triangle is oriented relative to the Sun. Observations of
the solar limb using the interferometer cannot possibly be made
accurate to 1 meter, and, even then, the distance from the solar
limb to the center of the Sun (for purposes of calculating the
solar potential) cannot possibly be modeled. One could conceive
of doing orbit determination using some typical Doppler or ranging
tracking of the spacecraft or even of using the precise relative
angular measurements from the interferometer themselves as a new
navigation data type.  However, experience with orbit
determination for free-flying interplanetary spacecraft shows that
non-gravitational forces do not allow anything like 1-meter
accuracy over periods of months.  If the impact parameter problem
is to be solved, then non-gravitational forces must be eliminated
using a drag-free system on board each spacecraft.

In our covariance study, we have assumed that the LATOR spacecraft
follow purely gravitational orbits (except for a constant
body-fixed acceleration that is solved for in the covariance study).  We
will find in our study that not only is the impact parameter
problem solved by the subsequent orbit determination, but that the
spacecraft-to-spacecraft distance problem is likewise resolved.  This is
to say that, as we simultaneously solve for orbit parameters for
both spacecraft along with the PPN parameters, any uncertainties
in the inertial positions of the spacecraft are correctly taken
into account through the correlations in the covariance matrix.
Thus, whatever final uncertainty we derive for the PPN parameters
will be the correct uncertainty, given the uncertainties in the
data-derived orbits of the spacecraft.

To determine how good a drag-free system needs to be to enable a
pure gravitational trajectory to be a good approximation, let us
consider the following:

\begin{enumerate}
\item A $10^{-13}$-radian angular accuracy at a distance of
$3\times10^{11}$ m corresponds to a position accuracy of 3 cm.
\item The position noise for a spacecraft suffering acceleration
noise with noise spectral density $S_a$ is $x_{\rm
rms}=\sqrt{S_a/(16\pi^4f^4)}$, at a Fourier frequency $f$. 
\item Therefore, if we require 3 cm accuracy at a frequency 1/month
($4\times10^{-7}$ Hz), we derive a drag-free requirement of
$S_a=3.6\times10^{-26}$ m$^2$s$^{-4}$Hz$^{-1}$, or a root spectral
density $\sqrt{S_a}=1.9\times10^{-13}$ ms$^{-2}$Hz$^{-1/2}$. 
\end{enumerate}
The currently planned LISA reference sensors are expected to have an accuracy of $3\times 10^{-15}\mathrm{m}\mathrm{s}^{-2}\mathrm{Hz}^{-1/2}$. Below $10^{-4} \mathrm{Hz}$, a $1/f$ increase is expected due to thermal drift resulting from the passive thermal isolation of the reference sensors. At $3\times 10^{-8} \mathrm{Hz}$, this would produce acceleration noise of $10^{-11} \mathrm{m}\mathrm{s}^{-2}\mathrm{Hz}^{-1/2}$, so a mK active thermal control (not a difficult technology) would be needed to keep thermal noise sources below the $2\times 10^{-13}\mathrm{m}\mathrm{s}^{-2}\mathrm{Hz}^{-1/2}$ we have assumed for the LATOR accelerometer. Voltage drifts in the measurement electronics do not affect the capacitance bridge used to read the accelerometer, since it is nearly a null measurement. Proof-mass charge control is absolute and independent of long-period effects. Thermal drift of the non-controlled parts of the spacecraft, along with mechanical relaxation of the spacecraft structure, {\it will} produce long-term self-gravity changes, but we have calculated these to be well out of the way of the level of performance we have assumed, as long as there is a level of spacecraft mechanical and thermal engineering comparable to that already required for LISA. Most other noise sources are white and LISA performance would keep them below the $10^{-15}\mathrm{m}\mathrm{s}^{-2}\mathrm{Hz}^{-1/2}$ LISA requirement. 

\section{Results and Conclusions}
The predicted accuracies of orbit parameters and spacecraft DC
accelerations from one year of LATOR data are shown in table
\ref{nuisres}.  The 60-cm uncertainty in the two semi-major axes
is primarily the result of strong correlation between these parameters.  (The uncertainty for $a_1$ or $a_2$ would have been about 60 microns
if one of these two parameters were being solved for alone.)
\begin{table}[h]
\begin{tabular}{|c|c|}
\hline
Parameter & LATOR predicted uncertainty \\
\hline
\hline
$a_1$, $a_2$ & 1.4 m \\
\hline
$e_1$, $e_2$ & $2 \times 10^{-11}$ \\
\hline
$\phi_{01}$, $\phi_{02}$ & $1.3 \times 10^{-10}$ \\
\hline
$\omega_1$, $\omega_2$ & $4 \times 10^{-11}$ \\
\hline
$a_{\perp 1}$,$a_{\perp 2}$ & $4 \times 10^{-14}$ $\textrm{m}/\textrm{s}^2$\\
\hline
$a_{\parallel 1}$,$a_{\parallel 2}$ & $3 \times 10^{-14}$ $\textrm{m}/\textrm{s}^2$ \\
\hline
\end{tabular}
\caption{\label{nuisres} Covariance study estimates for LATOR determination of
uncertainties in transmitter constants of the motion and DC
accelerations}
\end{table}

\begin{table}[h]
\begin{tabular}[c]{|c|p{5cm}|p{5cm}|p{4cm}|}
\hline
Parameter & Significance & Current best estimate & LATOR predicted uncertainty\\
\hline
\hline
$\gamma$ & Space-time curvature per unit rest mass. & $ 1 + (2.1 \pm 2.3)\times 10^{-5} $\cite{Bertotti:2003rm}. Equal to one in GR & $2.7\times 10^{-9}$\\
\hline
$\beta$ & Space curvature per unit gravitational self-energy & $1+(0.9\pm 1.1)\times 10^{-4}$\cite{2004PhRvL..93z1101W}. Equal to one in GR & Not solved for. \\
\hline
$\delta$ & Time curvature per unit gravitational self-energy & Unmeasured. Equal to one in GR & $1\times 10^{-3}$ \\
\hline
$J_2$ & Solar quadrupole moment parameter. & Umeasured. Estimate of about $10^{-7}$ inferred from solar models. & $5\times 10^{-9}$\\
\hline
\end{tabular}
\caption{\label{ppnres} Predicted uncertainties in PPN parameters with LATOR and
comparison with current best estimates.}
\end{table}

The results for the PPN parameters are shown in table
\ref{ppnres}. The uncertainty in $\gamma$ will represent a four
order-of-magnitude improvement over the current Cassini results
and will, for the first time, approach the level where heuristic
string-theory arguments predict that a deviation from GR might be
expected. The Parametrized Post-Post-Newtonian (PPPN) parameter $\delta$ will be measured for the first
time.  This is important because there are several theories whose
PPN parameters are identical with GR, but which differ from GR in
their predictions for the PPPN $\delta$ term.  Finally, we predict a 6\%
determination of the solar quadrupole moment, relative to its
expected value of $10^{-7}$.  This is a direct, model-independent
value and may be used as a touchstone value for various models of
the solar interior.

We conclude that LATOR will represent a significant step forward
in the experimental tests of relativistic theories of gravity and
will be able to satisfy the goals that were suggested in the
initial project proposals. We find that, with
a $1.9\times10^{-13}$ ms$^{-2}$Hz$^{-1/2}$ drag-free system covering a band down to nearly DC, it will be possible to simultaneously extract the PPN
parameters and all relevant orbit parameters from the
interferometer data alone.  This seems to be the only good
solution to the impact parameter problem, and we strongly suggest
that the engineering studies for LATOR consider this hardware
option.

\appendix
\section{Justification of Orbit Modelling}\label{orbjust}

The assumed accuracy of each interferometer measurement is $10^{-13}$ radians, which, at a distance of 2AU ($3\times 10^{11}$m) corresponds to 3 cm along-track position accuracy. As a comparison, let us consider the recent range data to various Mars spacecraft, the forces on Mars being purely gravitational.  The data for the last decade have post-fit residuals of 1-2 m.  The parameters that enter the solution include spacecraft locations and MarsÕ physical ephemeris for landers, orbital elements and MarsÕ gravity field for orbiters, corrections to Earth tracking station locations, solar corona and troposphere delays, orbits and masses of the planets, and the masses of several asteroids.  This combination of $\sim 200$ parameters in comparison with our 4 orbital elements would seem to suggest that we have oversimplified the problem.  However, the LATOR data and model are immune to the effects of most of these elements of the model.

The LATOR interferometers are in orbit above the troposphere.  The laser frequencies are high enough to avoid significant plasma propagation delays.  There is no local disturbing planet to complicate the spacecraft motion.  The interferometer makes an {\it angular} position measurement, so errors in the interferometer position cancel to first order.  Errors in the interferometer orientation likewise cancel when the differential position of the two spacecraft is calculated.  Finally, the long-period solar system model uncertainties in the position of each spacecraft are significantly reduced when the {\it differenced} spacecraft positions are calculated.  For an along-track sinusoidal error for each spacecraft of size $s$ and period $T$ (in days), the size of the error in the differenced position is given by
\begin{equation}\label{dperror}
\Delta s = \frac{s \cos(2\pi t/T)}{vT}\delta \lambda
\end{equation}
where $v$ is the velocity of the spacecraft in degrees per day and $\delta\lambda$ is the difference in heliocentric longitude between the two spacecraft, in degrees. Since $v\sim 1$ and $\delta\lambda\sim 2^{\circ}$, and since the solar system model errors are of order $\sim$years, the error in the differential position of the two spacecraft will be reduced by a factor $\frac{\delta \lambda}{T}$ (which is $\sim 0.01$ or less) relative to the single-spacecraft position error.  Thus a spacecraft error of 1-2 m becomes an error of 1-2 cm in the LATOR observable, slightly below the 3 cm error we assume in our covariance study.  It is therefore not an unreasonable assumption to treat the solar system parameters as known to this level, by virtue of the present and ongoing accurate Mars ranging data.

\begin{acknowledgments}
We would like to thank Slava Turyshev, Mike Shao and Ken Nordtvedt for help
understanding the LATOR design. We would also like to thank Slava Turyshev for reviewing our manuscript in advance of submission.
\end{acknowledgments}

\bibliography{LATOR_covariance_paper_revtex}
\end{document}